\documentclass[preprint, aps, pre, preprintnumbers,amsmath,amssymb,showpacs]{revtex4}

\usepackage[final]{graphicx} 

\newcommand{\bean}{\begin{eqnarray}}
\newcommand{\eean}{\end{eqnarray}}
\newcommand{\bea}{\begin{eqnarray*}}
\newcommand{\eea}{\end{eqnarray*}}
\newcommand{\beq}{\begin{equation}}
\newcommand{\eeq}{\end{equation}}

\def\vereq#1#2{\lower3pt\vbox{\baselineskip1.5pt \lineskip1.5pt
\ialign{$\hfill##\hfil$\crcr#2\crcr\sim\crcr}}}

\begin{document}

\title{Driving convection by a temperature gradient or a 
heat current}

\author{P.~Matura and M.~L\"{u}cke}
\affiliation{Institut f\"{u}r Theoretische Physik, Universit\"{a}t
des Saarlandes, D-66041~Saarbr\"{u}cken, Germany\\}

\date{\today}

\begin{abstract}
Bifurcation properties, stability behavior, dynamics, and the heat transfer of 
convection structures in a horizontal fluid layer that is driven away from 
thermal equilibrium by imposing a vertical temperature difference are compared 
with those resulting from imposing a heat current. In particular oscillatory 
convection that occurs in binary fluid mixtures in the form of travelling and
standing waves is determined numerically for the two different driving
mechanisms. Conditions are elucidated under which current driven convection 
is stable while temperature driven convection is unstable.
\end{abstract}


\maketitle

Many nonlinear dissipative systems that are driven away from 
thermal equilibrium show selforganization out of an unstructured state: A 
structured one appears when the driving exceeds a critical threshold 
\cite{CH93}. The driving might be realized by imposing a field 
gradient across the system --- e.g., 
a voltage difference across a semiconductor \cite{semicond} or a liquid crystal
\cite{liquid-crystal}, 
a temperature difference across a fluid layer \cite{BPA00}, or a 
concentration difference in a physical \cite{PCSN94}, chemical \cite{K03}, or 
biological \cite{CDFSTB03} system --- 
which drives a current. Or, alternatively, a current might be injected at 
one side of the system \cite{AhlersQT,WKPS85,MS86}. Now the question is, whether 
and how
the dissipative structures that form in response to these two different driving 
mechanisms are related to
each other concerning their dynamics, their structure, their stability behavior,
and their bifurcation properties.

We have investigated this question numerically for the case of convection
in a horizontal layer of a binary fluid like, e.g., ethanol-water 
\cite{early-exp}. Unlike one-component fluids like pure water, this 
system shows a surprisingly rich variety of different convection structures
already at
small driving \cite{CH93,PL84,LBBFHJ98}. There are spatially extended states of 
stationary convection
rolls and of temporally oscillating roll patterns in the form of traveling waves 
(TWs) and of standing
waves (SWs) that bifurcate out of the quiescent fluid state. In addition there
are also spatially localized traveling wave states that compete with extended 
convection structures. 
 

Here, we focus on convection in the form of straight
parallel rolls as they occur, e.g., in narrow channels with roll axes 
perpendicular to the long side walls. 
We have solved the appropriate hydrodynamic field equations 
\cite{PL84,LL87} with a finite-differences method \cite{BLKS95I} in a vertical 
$x-z$ cross section through the 
rolls perpendicular to their axes thus ignoring effects that come from field
variations along the roll axes \cite{PL84,CJT97,AlBa04}.  

Calculations were done for ethanol-water parameters, Lewis number
$L$=0.01 and Prandtl number $\sigma$=10. Results are presented here for two 
different separation ratios $\psi$=-0.03 and $\psi$=-0.1 \cite{coupling}
for which TW and SW solutions
bifurcate subcritically out of the quiescent fluid state via a common Hopf 
bifurcation. Our findings concerning the above posed questions are
representative also for TWs and SWs at stronger, i.e., more negative Soret 
coupling strength $\psi$. 
 

The horizontal boundaries at top ($z$=1) and bottom ($z$=0) are no-slip, 
impermeable, and perfectly heat conducting, thus enforcing the absence of 
lateral temperature gradients there. Sapphire or copper plates provide a good 
experimental approximation. 
Two different experimentally realizable horizontal boundary conditions (bc) for
the temperature are explored here: {\em (i)} Dirichlet bc of fixed temperatures
(constant in space and time) at $z$=0 and $z$=1 with a difference of $\Delta T$ 
between them and {\em (ii)} von Neumann bc of fixed total vertical heat current
at $z$=0 and Dirichlet bc of fixed temperature at $z$=1. 
At the impermeable boundaries the vertical concentration 
current vanishes and consequently the local vertical heat current reduces there 
to $-\partial_z T$ \cite{LL87}. Note that we impose in case {\em (ii)} the
horizontal mean
\begin{equation}
Q=-\overline{\partial_z T|_{z=0}}
\end{equation}
of the heat current at the lower side of the fluid layer so that the total heat
current injected into it is a constant. We shall identify the driving 
conditions of case {\em (i)} by TT for short and those of case {\em (ii)} by QT.

Laterally we impose for all fields periodic bc 
with wavelength $\lambda$=2. This is roughly the critical one for
onset of oscillatory convection. Moreover, it is often seen also in 
nonlinear convection with rolls of about circular shape. Finally, to determine 
SW solutions that are unstable against
horizontal mirror symmetry breaking phase propagation we enforce horizontal 
mirror symmetry, say, at $x$=0 thereby fixing the phase \cite{MDL04}. 

As control parameter measuring the strength of the driving we use in the TT 
case the relative deviation  
\begin{equation}
\epsilon=\Delta T/\Delta T_c -1
\end{equation}
from the critical temperature difference $\Delta T_c$ for onset of convection.
The driving in case QT is measured by the relative deviation 
\begin{equation}
\rho= Q/Q_c -1
\end{equation}  
from the critical imposed heat current.

\begin{figure}
\centerline{\includegraphics[width=12cm]{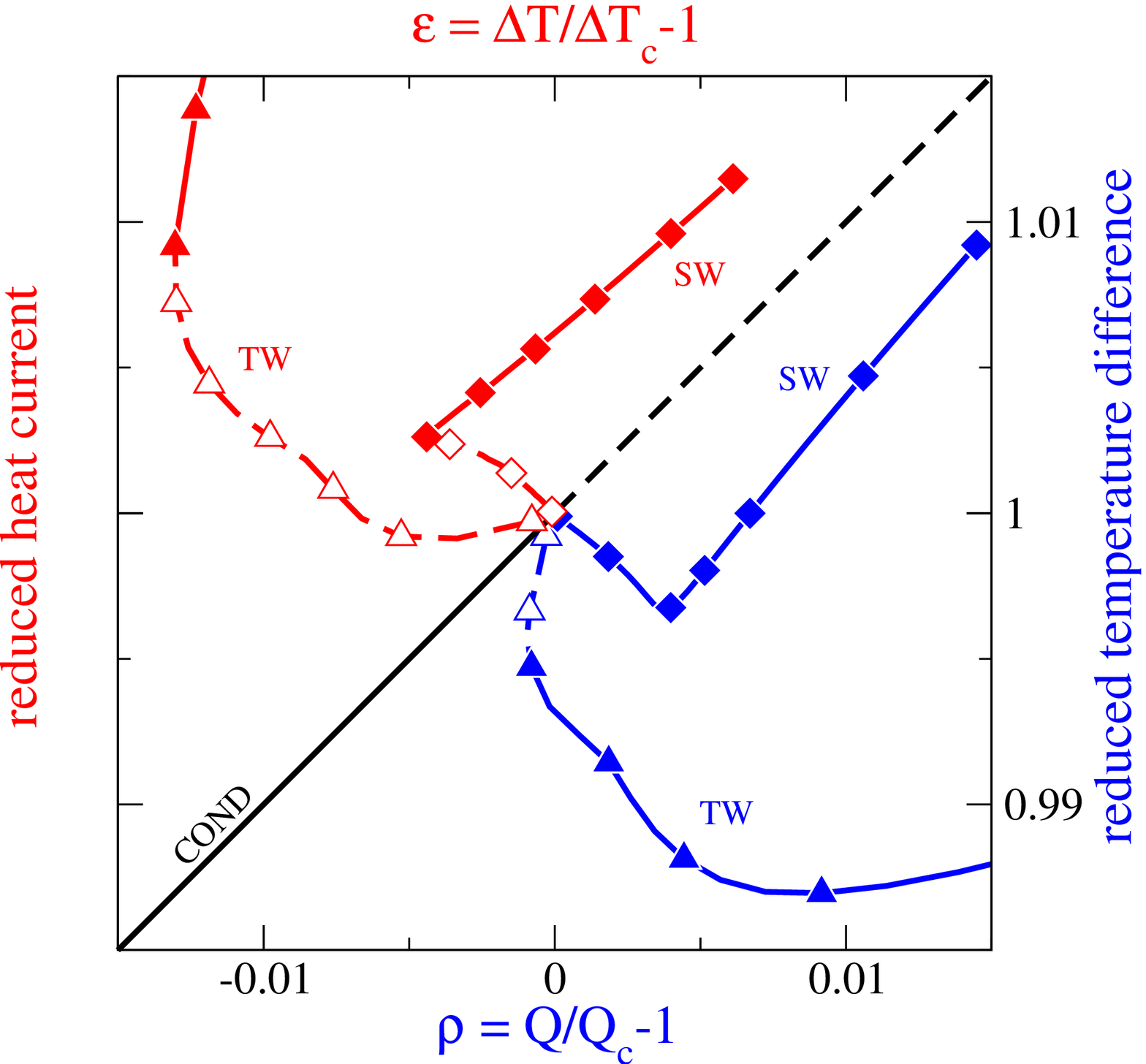}}
\caption{(color online) Bifurcation diagrams of heat current and temperature 
difference for
oscillatory convection at $\psi=-0.03$ subject to different bc. The upper left
(lower right) part shows $Q/Q_c$ ($\Delta T/\Delta T_{c}$) for TT (QT) driving
versus $\epsilon$ ($\rho$) on the upper (lower) abscissa.
For SWs the time averages $\left<Q\right>/Q_c$ and $\left<\Delta T\right>/\Delta T_{c}$, respectively,
are plotted. The bisecting line marks the quiescent conductive state.
Full (dashed) lines and filled (open) symbols denote stable (unstable) states. 
SW solutions were obtained with phase pinning conditions; otherwise they are
completely unstable.} 
\label{FIG:QT003}
\end{figure}
\begin{figure}
\centerline{\includegraphics[width=12cm]{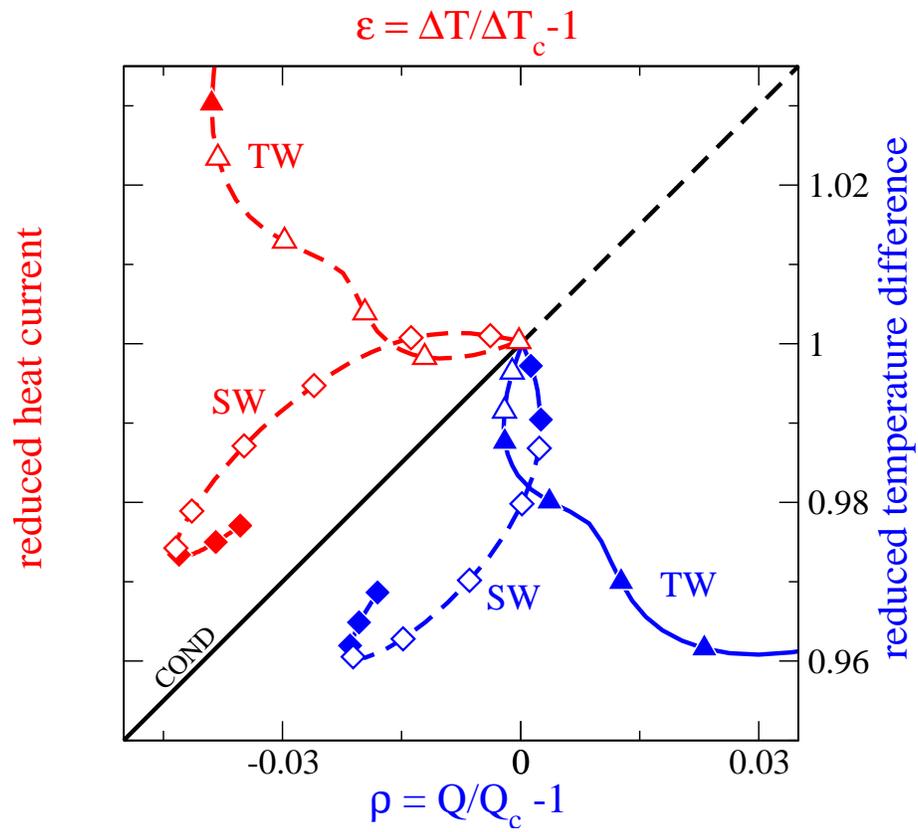}}
\caption{(color online) Bifurcation diagrams of heat current and temperature 
difference as in Fig.~\ref{FIG:QT003}. Here, however, for $\psi=-0.1$.} 
\label{FIG:QT01}
\end{figure}

We shall discuss first TW convection and then SW solutions. In 
Figs.~\ref{FIG:QT003} and \ref{FIG:QT01} we show for two different $\psi$ the
bifurcation diagrams of nonlinear relaxed TW states: {\em (i)} current $Q$ as a
function of $\epsilon$ for TT driving and {\em (ii)} temperature difference 
$\Delta T$ versus $\rho$ for the QT case. Note that $Q$ as well as $\Delta T$ 
are constant for TWs. The diagonal line shows the linear diffusive relation
$Q_{cond}/Q_c=\Delta T_{cond}/\Delta T_c$ of the quiescent conductive state. It 
loses
stability via a Hopf bifurcation at $\Delta T_c$ or $Q_c$, respectively.
After increasing the driving slightly beyond this threshold transient growth of
oscillatory convection occurs with increasing $Q$ for the TT case. For QT
conditions $\Delta T$ decreases since convection cools the lower boundary.
Initially, the oscillations have the Hopf frequency. But finally, the TT 
transient ends in a stationary convection state 
since the TW branch terminates with zero frequency in a stationary 
solution branch already below $\epsilon$=0 for our $\psi$'s \cite{rstar}. On 
the other hand, the QT growth transient ends in a relaxed nonlinear TW 
(lower part of Figs.~\ref{FIG:QT003} and \ref{FIG:QT01}). 

The curves of $Q/Q_c$ versus $\epsilon$ and of $\Delta T/\Delta T_c$ versus 
$\rho$ in Figs.~\ref{FIG:QT003} and \ref{FIG:QT01} are reflections of each
other at the diagonal, bisecting conduction line. Note, however, that the 
transients and the stability ranges of the relaxed TWs are different. Concerning
the latter, for example, the hysteresis interval
in $\rho$ for QT is significantly smaller than the one in $\epsilon$ for TT
since a large portion of unstable TT generated TWs below onset gets stabilized
under QT driving. 

\begin{figure}
\centerline{\includegraphics[width=12cm]{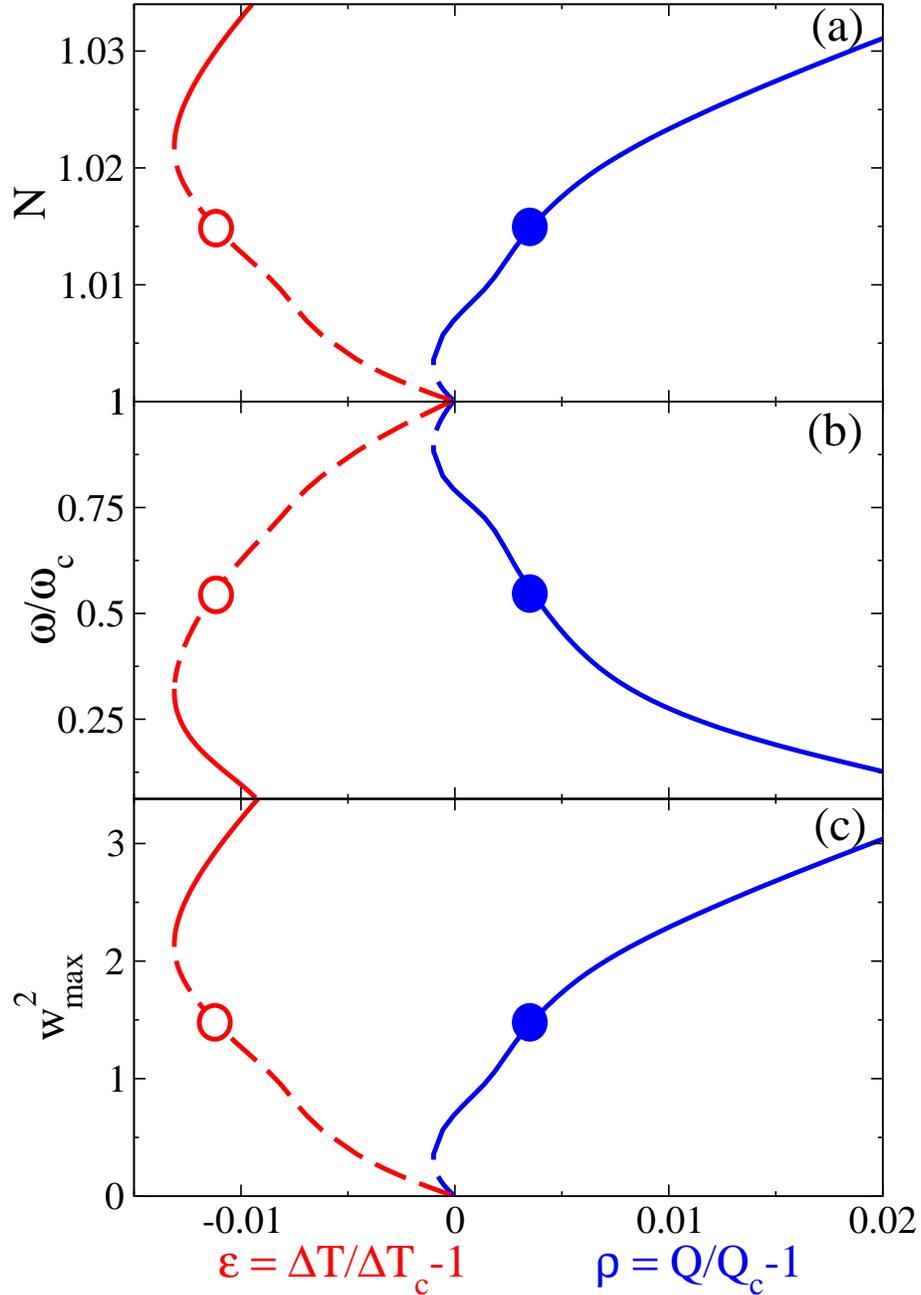}}
\caption{(color online) TW bifurcation diagrams: (a) Nusselt number $N$, 
(b) reduced oscillation frequency $\omega/\omega_{c}$, and (c) squared maximal 
vertical velocity $w_{max}^{2}$ for the TWs of Fig.~\ref{FIG:QT003} with 
$\psi=-0.03$. The left and right curves refer to TT and QT driving, 
respectively. Full (dashed) lines denote  stable (unstable) TWs. Symbols 
identify examples of equivalent TWs for equivalent control parameters 
$\epsilon$ and $\rho$.} 
\label{FIG:TW003}
\end{figure}

In Fig.~\ref{FIG:TW003} we show for the TWs of Fig.~\ref{FIG:QT003} bifurcation
diagrams of Nusselt number $N$, reduced frequency $\omega/\omega_{c}$,
and squared maximal vertical velocity $w_{max}^{2}$ versus the respective
control parameters. The  Nusselt number
\begin{equation} 
N=Q/Q_{cond}=(Q/Q_c)\Delta T_c/\Delta T 
\end{equation}
provides the relation 
\begin{equation} \label{EQ:rho-eps}
\rho = (1+\epsilon)N -1
\end{equation}
between equivalent control parameters $\epsilon$ and $\rho$ corresponding to
reflection at the conductive diagonal in Figs.~\ref{FIG:QT003} and 
\ref{FIG:QT01}: 
TWs that are generated by TT or QT driving at $\epsilon$- and $\rho$-values
related by (\ref{EQ:rho-eps}) have the same spatiotemporal properties, e.g., the 
same $N,\omega,w_{max}$ as indicated by the symbols in Fig.~\ref{FIG:TW003}. 
Their stability, however, might differ. 

Eq.~(\ref{EQ:rho-eps}) yields also the relation
\begin{equation}\label{EQ:slope-relations} 
\partial_{\rho}A=\partial_{\epsilon}A/[N+(1+\epsilon)\partial_{\epsilon}N]
\end{equation}
between the slopes $\partial_{\rho}A(\rho)$ and $\partial_{\epsilon}A(\epsilon)$
in the QT and TT bifurcation diagrams of any order parameter $A$ (say,
$N,\omega,w_{max}^2, etc)$ versus $\rho$ or $\epsilon$, respectively. Hence, the
QT bifurcation becomes already tricritical, i.e., it changes from backwards to 
forwards
when the initial slope $s=\partial_{\epsilon}N(\epsilon=0)$ of the TT Nusselt 
number increases beyond $-1$. In other words, all TT driven backwards
bifurcating unstable TWs for which $s>-1$ can be stabilized by switching over 
to QT driving. 

Note that the relations (\ref{EQ:rho-eps}) and 
(\ref{EQ:slope-relations}) hold also for
any stationary convection solution so that the bifurcation diagrams of  
$Q/Q_c$ versus $\epsilon$ and of $\Delta T/\Delta T_c$ versus $\rho$ are 
reflections of each other. Thus, the 
stabilization effect of QT driving holds also for any
stationary convection that bifurcates backwards with TT \cite{Busse}. The 
(stability)
properties of forward bifurcating stationary convection remain unchanged 
when using QT instead of TT conditions.

\begin{figure}
\centerline{\includegraphics[width=12cm]{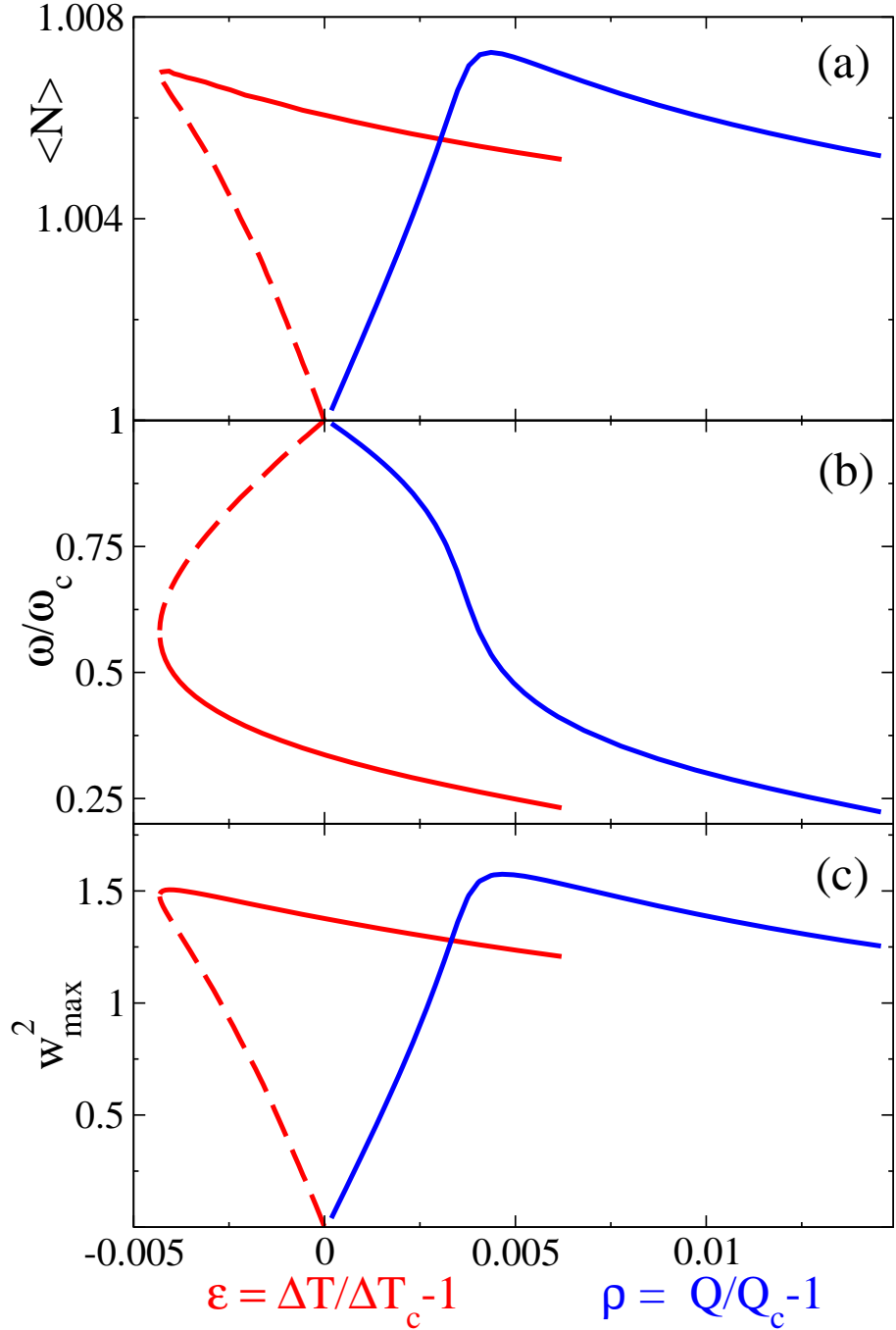}}
\caption{(color online) SW bifurcation diagrams: 
(a) time averaged Nusselt number $\left<N\right>$, 
(b) reduced oscillation frequency $\omega/\omega_{c}$, and (c) squared maximal 
vertical velocity $w_{max}^{2}$ for the SWs of Fig.~\ref{FIG:QT003} with 
$\psi=-0.03$. The left and right curves refer to TT and QT driving, 
respectively. Full (dashed) lines denote  stable (unstable) SWs.} 
\label{FIG:SW003}
\end{figure}
\begin{figure}
\centerline{\includegraphics[width=12cm]{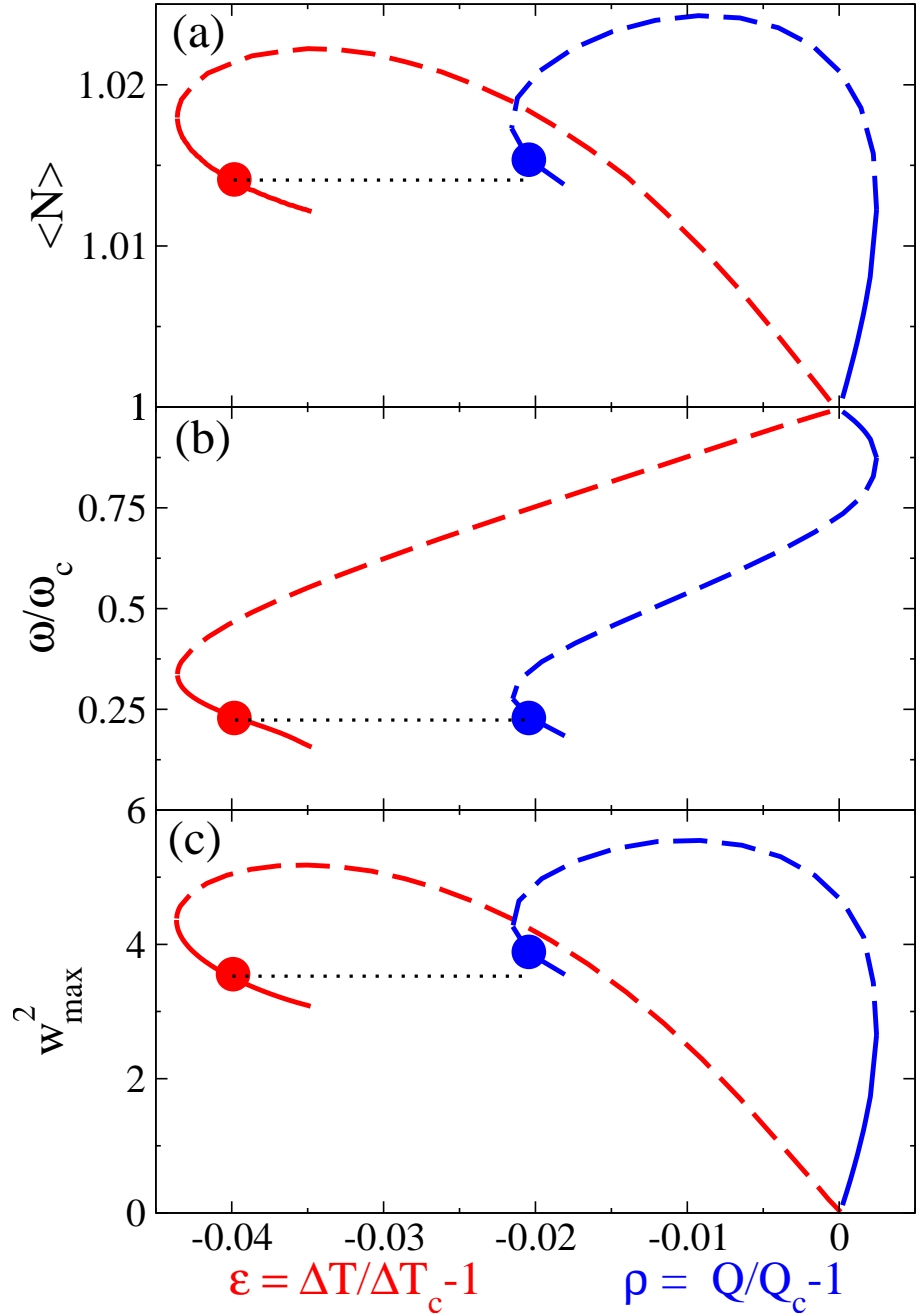}}
\caption{(color online) SW bifurcation diagrams as in Fig.~\ref{FIG:SW003} but 
for the SWs of Fig.~\ref{FIG:QT01} with $\psi=-0.1$. Symbols identify examples 
of SWs with the same frequency.} 
\label{FIG:SW01}
\end{figure}
\begin{figure}
\centerline{\includegraphics[width=12cm]{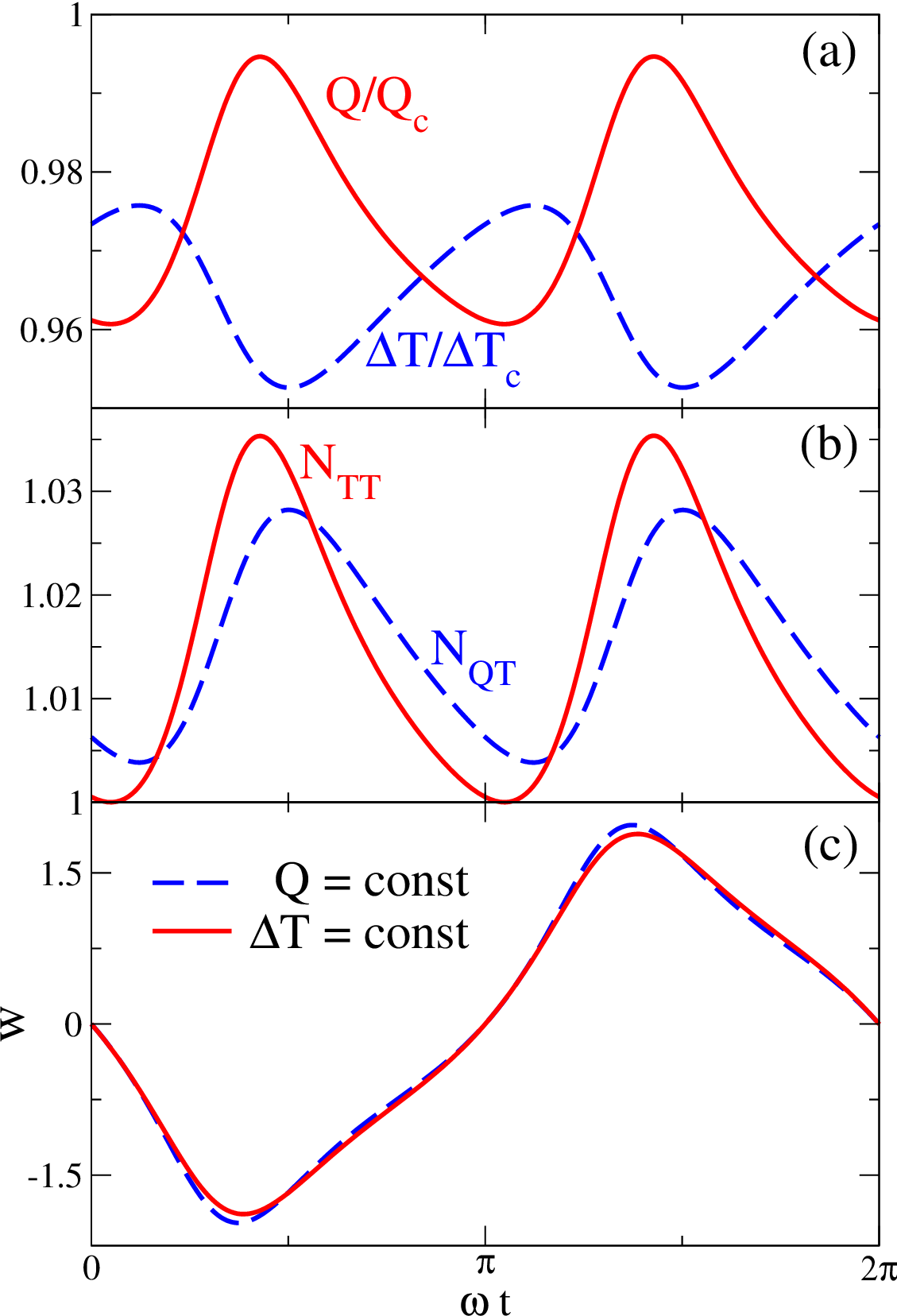}}
\caption{(color online) Oscillation profiles of the SWs marked in 
Fig.~\ref{FIG:SW01} by symbols. (a) Current and temperature oscillation  
subject to TT and QT driving, respectively. (b) Nusselt numbers
$N_{TT}(t)= [Q(t)/Q_c]\Delta T_c/\Delta T = Q(t)/Q_{cond}$ and 
$N_{QT}(t)= (Q/Q_c)[\Delta T_c/\Delta T(t)] =\Delta T _{cond}/\Delta T(t)$. 
(c) Vertical velocity $w$ at midheight between two rolls.} 
\label{FIG:VglSW}
\end{figure}

In the remainder of this letter we dicuss SW convection. Under TT (QT) driving 
the heat current $Q$ (temperature difference $\Delta T$) oscillates with twice 
the SW frequency \cite{SWoscillations}. So, in Figs.~\ref{FIG:QT003} and 
\ref{FIG:QT01} we show bifurcation diagrams of the time averages $\left<Q\right>/Q_c$ and 
$\left<\Delta T\right>/\Delta T_{c}$, respectively. Like for TWs, QT conditions have a 
stabilizing effect also on SWs. Note, however, that the two SW solution branches
in these Figs. are not reflections of each other  
at the conduction diagonal. Their spatiotemporal properties differ and the 
relation $\left<\rho\right> = (1+\epsilon)\left<N\right> -1$ provides only an approximate 
equivalence between the bifurcation diagrams of
the order parameters in Fig.~\ref{FIG:SW003} and \ref{FIG:SW01}. For example, 
the SWs marked by symbols in Fig.~\ref{FIG:SW01} have the same frequency. But
$w_{max}$ differs slightly and so does $\left<N\right>$ --- i.e., 
$\left<Q/Q_c\right>\Delta T_c/\Delta T$ for TT driving in comparison with 
$(Q/Q_c)\left<\Delta T_c/\Delta T\right>$ for QT driving. Also the oscillations of the 
flow differ slightly [Fig.~\ref{FIG:VglSW}(c)]. On the other hand, the 
oscillation profile of $Q(t)$ differs significantly from 
the one of $\Delta T(t)$ [Fig.~\ref{FIG:VglSW}(a)] and also the profile of 
$N_{TT}(t)$ differs from the one of $N_{QT}(t)$ [Fig.~\ref{FIG:VglSW}(b)].

In summary: Driving convection with a fixed heat current can stabilize
SW, TW, and stationary states that bifurcate backwards and that are unstable 
with imposed temperature difference. However, irrespective of their stability 
relaxed TWs for the two driving mechanisms are simply related to each other. 
The same holds for stationary solutions. But the time evolution
of, say, growth transients differ in general. Also SW oscillations driven by a
constant field gradient differ from those resulting from constant current
driving. It would be interesting to see how far these different conditions 
influence the spatio temporal properties of convection structures with more complex 
dynamics.


\clearpage
\clearpage
\clearpage
\clearpage
\clearpage
\clearpage


\begin{thebibliography}{999}
\bibitem{early-exp}
Early experiments done with an isopropanol-water mixture
in two different set-ups showed different hysteresis loops in
Schmidt-Milverton plots of temperature difference (on the ordinate) versus 
heat current (on the abscissa) depending on which of the two was imposed 
\cite{VP84}.
\bibitem{coupling}
For 1.4 weight \% of ethanol mixed into water at $T=20^{\circ} C$ the 
separation ratio measuring the Soret coupling strength between temperature and
concentration gradients is $\psi \simeq -0.1$ \cite{KWM88,CJT97}.
\bibitem{Busse}
Pure fluids with temperature dependent material parameters that are showing a 
stationary backwards bifurcation for TT driving were predicted \cite{B67} 
to allow oscillations under QT conditions. But we did not find such a behavior 
in our system.
\end{thebibliography}
\end{document}